\documentclass[aps,prl,showpacs,amsmath,twocolumn,amssymb,superscriptaddress,letterpaper]{revtex4}
\usepackage{graphicx,color}
\usepackage{amssymb}   
\usepackage{amsmath}
\usepackage{epstopdf}
\usepackage{natbib}
\usepackage{hyperref}
\usepackage{bm}
\usepackage{color}

\begin{document}
\title{Non-Abelian statistics of Majorana zero modes in the presence of an Andreev bound state}
\author{Wenqin Chen}
\affiliation{Department of Applied Physics, School of Science, Xian Jiaotong University, Xian 710049, China}
\author{Jiachen Wang}
\affiliation{Department of Applied Physics, School of Science, Xian Jiaotong University, Xian 710049, China}
\author{Yijia Wu}
\affiliation{International Center for Quantum Materials, School of Physics, Peking University, Beijing 100871, China}
\author{Jie Liu}
\email{jieliuphy@xjtu.edu.cn}
\affiliation{Department of Applied Physics, School of Science, Xian Jiaotong University, Xian 710049, China}
\author{X. C. Xie}
\affiliation{International Center for Quantum Materials, School of Physics, Peking University, Beijing 100871, China}
\affiliation{Beijing Academy of Quantum Information Sciences, Beijing 100193, China}
\affiliation{CAS Center for Excellence in Topological Quantum Computation,
University of Chinese Academy of Sciences, Beijing 100190, China}

\begin{abstract}
Braiding Majorana zero modes (MZMs) is the key procedure toward topological quantum computation. We show such braiding can be well performed in a parallel semiconductor-superconductor nanowire structure. Considering the fact that the low-energy Andreev bound states (ABSs) usually mix with the MZMs in the present set-up, we further investigate the braiding properties of MZMs when an ABS is presented. Our numerical simulation suggests that ABS can be regarded as a pair of weakly coupled MZMs. The dynamical hybridization of MZMs plus the non-Abelian braiding of MZMs would induce an arbitrary rotation on the Bloch sphere of a single qubit. Remarkably, such rotation is manipulable since the rotation parameters could be individually modulated. Thus, the dynamic evolution can be eliminated and the non-Abelian braiding statistics, independent of the braiding time, retrieves.
\end{abstract}

\maketitle

{\emph {Introduction}} --- Majorana zero mode (MZM) is deemed as the most promising candidate for topological quantum computation (TQC) \cite{kitaev, nayak} for its non-Abelian statistics. The exploration for MZMs in topological superconductors (TSCs) has been drawing extensive attention in the last decade \cite{Fu, sau, fujimoto, sato, alicea2, lut, oreg, potter, 2DEG1, 2DEG2}.
To date, TSC has been realized in various experimental platforms \cite{kou, deng, das1, hao1, hao2, Marcus, perge, Yaz2, Jia, Fes1, Fes2, Fes3, PJJ1, PJJ2}. The semiconductor- superconductor heterostructure, first experimentally realized one among these systems, is regarded as one of the most promising platforms to realize TQC. However, in spite of the promising signs, other complications such as the Andreev bound states (ABSs) still cannot be ruled out \cite{Jie1, brouwer, Aguado, ChunXiao1, Moore, Wimmer, Aguado2, Klinovaja, Tewari, ABS, ABSM}. The ABS, which is viewed as a pair of ``fake'' MZMs, is widely observed in experiments and hard to get rid off in the present set-up \cite{ChunXiao1, Wimmer, Aguado2, ABSM}.
Though various experimental schemes have been proposed \cite{ChunXiao1, Moore, Wimmer, Aguado2, ABS1, ABS2, ABS3, ABSM1, ABS4, ABS5, ABS6, ABS7, ABS8, ABS9, ABS10,ABS11,ABS12}, there is still no convincing way to completely distinguish these two types of states.

\begin{figure}
\centering
\includegraphics[width=3.25in]{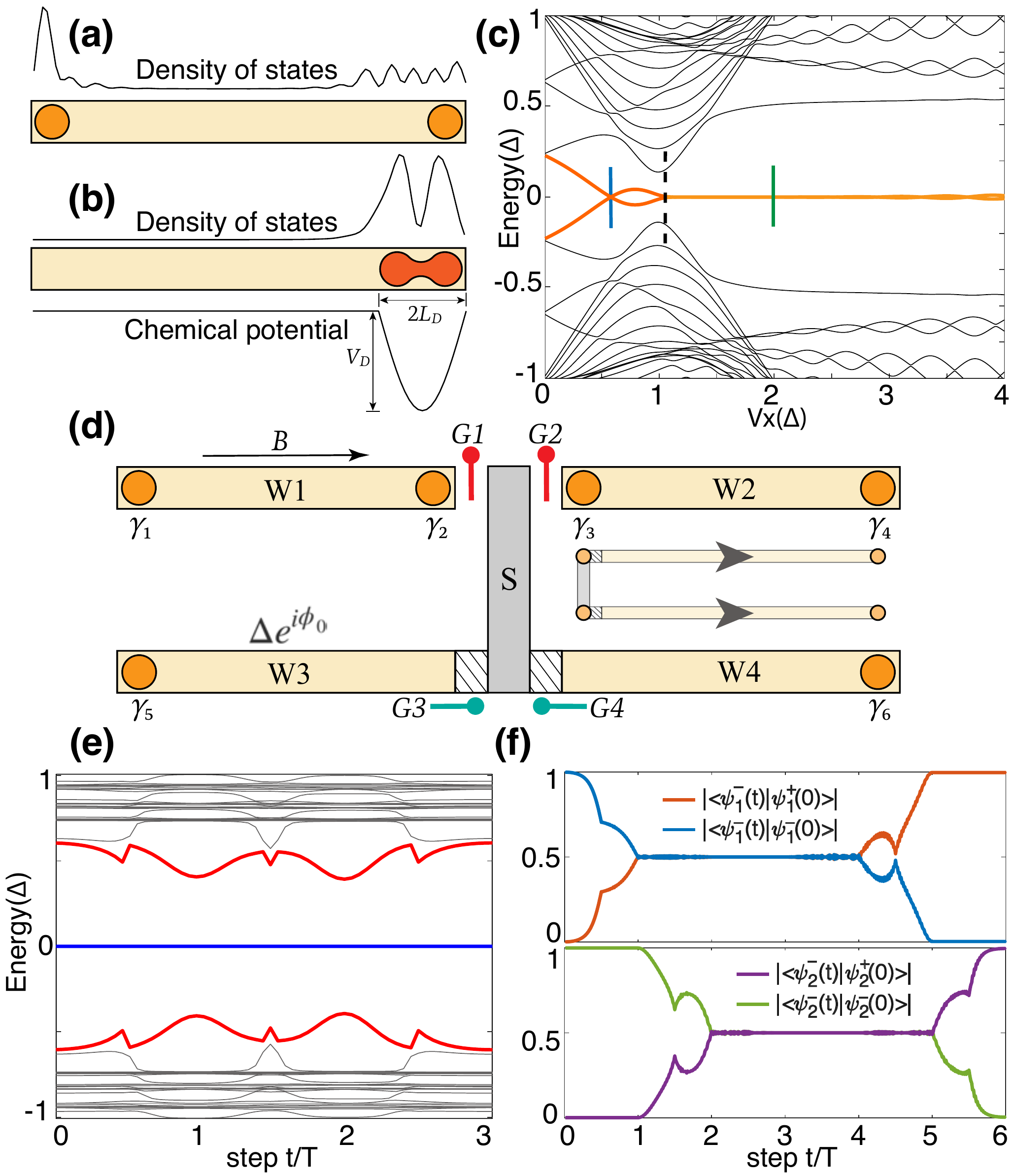}
\caption{
(a) MZMs in a semiconductor-superconductor nanowire.
(b) ABS in the same nanowire with a QD confinement presented.
(c) The energy spectrum of the semiconductor-superconductor nanowire (with the QD confinement) versus the Zeeman energy.
(d) The modified ``tetron'' structure adopted for the braiding of the MZMs, the bottom nanowires ($W3$, $W4$) are ancillary ones during the braiding. All four arms are topologically non-trivial with $N_x=100a$, $\mu=-2t_0$, and $V_x = 2\Delta$ [vertical green line in (c)]. The central region(S) is a trivial superconductor with length $N_c = 5a$, and other parameters are the same as those in the nanowires.
(e) The energy spectrum of the system during the braiding with $\phi_0=0.5\pi$, in which the gap is always open.
(f) Evolution of the wavefunction $\psi_j^{-}(t)$ during the braiding.
}
\label{f1}
\end{figure}

The most likely way of distinguishing MZM from the ABS is certainly based on its non-Abelian statistics \cite{Ivanov, alicea3,NQP, TQC1}.
However, only few studies have paid attention to this topic thus far \cite{TQC2,TQC3,TQC4}. The main obstruction lies in two aspects. First, the braiding protocols which have been proposed so far are quite complicated and hard to realize experimentally \cite{alicea3, NQP}.
Recently, a parallel structure is deemed as the most feasible way to achieve braiding operation \cite{MSQ,net1,net2,Yu1}. However, the projective-measurement-based braiding in such structure is not under control due to its probabilistic nature.
Second, since the ABSs are hard to get rid off with the state of art nanotechnology, it is necessary to study the braiding statistics in the presence of both MZM and ABS. Such investigation will also shed light on how to realize or modify TQC when ABS is engaged.

This Letter is motivated by attempting to solve the above two problems. First, we put forward a modified parallel structure [Fig. \ref{f1}(d)] and show that the non-Abelian braiding of MZMs can be conveniently performed in such structure in a definite fashion. Second, we clarify the MZMs' braiding rule in the condition that one pair of MZMs is replaced by an ABS. Our braiding results suggest that an ABS can be decomposed into two weakly coupled MZMs. By combining the hybridization-induced evolution of the ABS and the non-Abelian braiding of the MZMs, an rotation to arbitrary point on the single-qubit's Bloch sphere can be implemented.
Such rotation is manipulable since each parameter can be individually modulated in the corresponding braiding section. In this way, the dynamic evolution can be eliminated and the non-Abelian braiding statistics, independent of the braiding time, retrieves. We conclude that though the presence of ABS creates complication, the non-Abelian braiding properties of the MZMs can still be observed.

{\emph {Modified ``tetron'' structure for non-Abelian braiding}} ---Since the magnetic field can only be oriented along one direction, a parallel aligned structure ``tetron'' qubit is proposed for realizing the braiding through a series of projective measurements \cite{TQC2, net1, net2, MSQ}. The ``tetron'' qubit consists of two parallel nanowires which are connected through a trivial superconductor. Hence the ``tetron'' qubit can be described by the Hamiltonian as $H_{\mathrm{T}} = \sum_iH_{i} + H_{\mathrm{S}} + H_{\mathrm{Tc}}$. Here $H_{i}$ is the Hamiltonian for the $i$-th nanowire whose explicit form is:

\begin{eqnarray}
\label{model}
 H_{i} &=& \sum\nolimits_{\mathbf{R},\mathbf{d},\alpha} { - t_0(\psi _{\mathbf{R} + \mathbf{d},\alpha }^\dag \psi_{\mathbf{R},\alpha } + h.c.) - \mu \psi_{\mathbf{R},\alpha }^\dag \psi_{\mathbf{R},\alpha } } \nonumber \\
&+& \sum\nolimits_{\mathbf{R},\mathbf{d},\alpha ,\beta } { - i{U _R} \psi _{\mathbf{R} + \mathbf{d},\alpha }^\dag  \hat z \cdot (\vec{\sigma}  \times \mathbf{d})_{\alpha \beta }   \psi _{\mathbf{R},\beta } } \nonumber \\
&+& \sum\nolimits_{\mathbf{R},\alpha} \Delta e^{i\phi} \psi _{\mathbf{R},\alpha }^{\dagger} \psi _{\mathbf{R},-\alpha }^{\dagger} + h.c. \nonumber \\
&+& \sum\nolimits_{\mathbf{R},\alpha ,\beta } { \psi _{\mathbf{R}, \alpha }^\dag (V_x\vec{\sigma}_x)_{\alpha \beta} \psi _{\mathbf{R}, \beta}}.
\end{eqnarray}

\noindent where $\mathbf{R}$ denotes the lattice site, $\mathbf{d}$ is the unit vector, $\alpha$ and $\beta$ are the spin indices, $t_0$ denotes the hopping amplitude, $\mu$ is the chemical potential, $U_{R}$ is the Rashba coupling strength, and $V_x$ is the Zeeman energy. The superconducting pairing amplitude and the pairing phase are denoted as $\Delta$ and $\phi$, respectively. Besides, $H_{\mathrm{S}}$ corresponds to the trivial superconductor which has the same form as $H_i$ except for $U_{R}=0$. $H_{\mathrm{Tc}}$ is the coupling term connecting the trivial superconductor and the four nanowires. The parameters adopted here are specified referring to the experiment \cite{kou} as $\Delta=250 \mu \text{e}V$, $t_0=10\Delta$, and $U_{R}=2\Delta$.

To remove the probabilistic nature of the projective measurement in the original tetron stucture, we propose a modified tetron structure as illustrated in Fig. \ref{f1}(d), in which four additional gates ($G1\sim G4$) are located near the intersection of the nanowires and the trivial superconductor.
In such structure, the braiding can be realized in a definite way through tuning the gate voltages in the corresponding gates\cite{Sato,Roy}.
Before the braiding, all four arms ($W1\sim W4$) are topologically non-trivial, and gate voltages in $G1$ and $G2$ are turned on while in $G3$ and $G4$ are turned off, hence three pairs of MZMs ($\gamma_{2j-1}$ and $\gamma_{2j}$, $j = 1,2,3$) are localized at the ends of the three divided sections.
The braiding protocol takes three steps (the time-cost for each step is $T$) to swap $\gamma_2$ and $\gamma_3$ spatially. In step 1, $G1$ is turned off and then $G3$ is turned on, hence $\gamma_2$ is teleported to $W3$. In step 2, $G2$ is turned off and then $G1$ is turned on, so $\gamma_3$ is teleported to the original position of $\gamma_2$. In step 3, $G3$ in turned off and then $G2$ is turned on. Consequently, the spatial positions of $\gamma_2$ and $\gamma_3$ are swapped.

 To obtain the correct braiding results, the topological gap is required to remain open during the braiding. Therefore, a finite superconducting phase difference $\phi_0$ should be kept between the top and the bottom nanowires as shown below. If $\phi_0=0$, a domain wall (DM) structure \cite{Spin} is formed when only $G2$ and $G3$ are turned off [the magnetic field are pointed along one direction so that the center region is twisted, see the inset of Fig. \ref{f1}(d)].
In such case, an additional pair of end modes which emerges from the bulk and collapses at the zero energy will ruin the braiding. On the contrary, when $\phi_0\neq0$, the DW structure is smoothed out so that the gap remains open during the braiding [e.g. $\phi_0=0.5\pi$, see Fig. \ref{f1}(e)] and the braiding results remain valid. Besides, the topological gap decreases exponentially with the increase of the length of the trivial superconductor $N_c$. Hence, it is better to choose $N_c$ in the same order as the MZMs' coherence length. The latter is usually in the order of $10^{2}$ nanometers and such length scale is feasible in the state-of-art technology.

\begin{figure}
\centering
\includegraphics[width=3.25in]{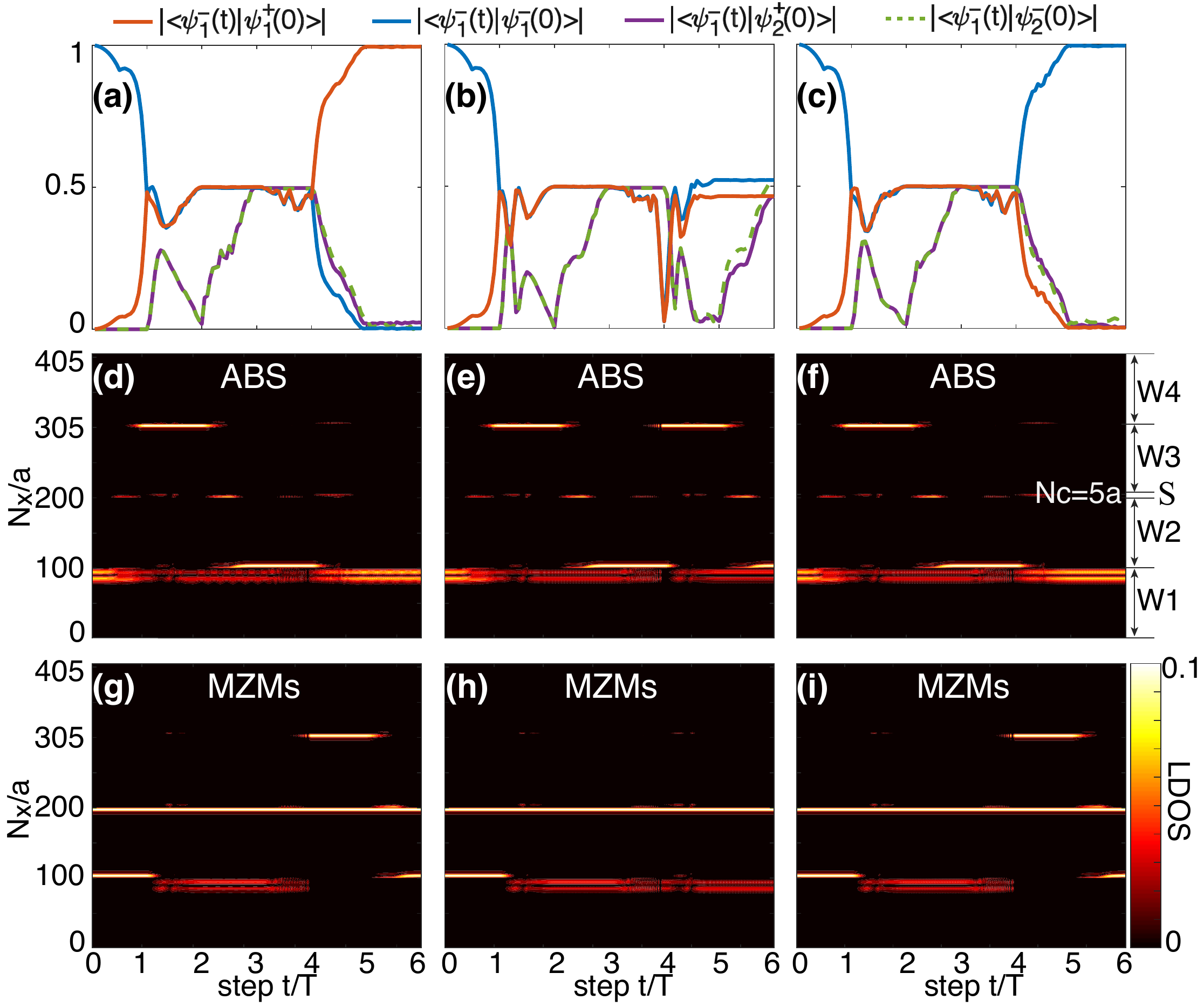}
\caption{
Evolution of the wave function $\psi_1^{-}=(\gamma_1-i\gamma_2)/\sqrt{2}$ with braiding time-cost (a) $T = 400/\Delta$; (b) $T = 450/\Delta$; and (c) $T = 500/\Delta$.
Here the Zeeman energy in the left arm {\color{red}($W1$)} is $V_x=0.6\Delta$, hence the left arm is in the trivial regime with an ABS presented [see the blue vertical line in Fig. \ref{f1}(c)]. All other parameters are the same as those in Fig. \ref{f1}(d).
(d)-(f) Evolution of the LDOS for ABS. The parameters correspond to (a)-(c), respectively. The MZMs are splitted and fused as indicated by the LDOS.
(g)-(i) Corresponding evolution of the LDOS for MZMs.
}
\label{f2}
\end{figure}

With the topological gap kept open, our numerical simulation [Fig. \ref{f1}(f)] confirms that the non-Abelian braiding can be accomplished in such structure. The braiding of MZMs can be represented by the operator $B(\gamma_i,\gamma_j) = \exp(\frac{\pi}{4}\gamma_i\gamma_j)$, which transform the MZMs as $\gamma_i\rightarrow \gamma_j$ and $\gamma_j\rightarrow -\gamma_i$ \cite{Ivanov}.
The effective low-energy Hamiltonian describing the MZMs in each separated part of the ``tetron'' before braiding is in the form of $H_{j, \mathrm{eff}} = i \epsilon_{j} \gamma_{2j-1} \gamma_{2j}$ ($j = 1,2,3$). Hence the eigenstates are in the wavefunctions of $\psi_j^{\pm}(0) = (\gamma_{2j-1}\pm i\gamma_{2j})/\sqrt{2}$.
If $\gamma_2$ and $\gamma_3$ are swapped twice in succession, then the wavefunction will evolve into $\psi_1^{\pm} (6T) =  (\gamma_1\mp i\gamma_2)/\sqrt{2}=\psi_1^{\mp}(0)$ and $\psi_2^{\pm} (6T) =  (-\gamma_3\pm i\gamma_4)/\sqrt{2}=-\psi_2^{\mp}(0)$.
 We simulate the wavefunction evolution during the braiding as $|\psi_j^{\pm}(t)\rangle=U(t)|\psi_j^{\pm}(0)\rangle$, where $U(t) = \hat{T} \exp[i\int_0^{t}d\tau H(\tau)]$ is the time-evolution operator and $\hat{T}$ is the time-ordering operator \cite{Jie2, Jie3,Dirac}. The simulation results confirm that $\psi_{j}^{+}$ evolves into $\psi_{j}^{-}$ ($j = 1,2$) after adiabatically swapping $\gamma_2$ and $\gamma_3$ twice in succession, which demonstrate the non-Abelian braiding rules discussed above.

{\emph {Non-Abelian braiding in the presence of ABS}} --- We further investigate the braiding rule in the presence of an ABS. ABS is usually induced by the inhomogeneity at the interface, which can be modeled by a quantum dot (QD) confinement at the end of the nanowire \cite{Marcus, ChunXiao1, ABSM}.
As depicted in Fig. \ref{f1}(b), a sinusoidal local chemical potential in the form of $V_{d}(R)=-V_D\cos(2\pi\frac{R-L_D}{2L_D})$ is presented at the right end of the nanowire, where $L_D=10a$ is the half width of the QD and $V_D=0.1t_0$ is the depth of the potential well.
When an external magnetic field is applied, as shown in Fig. \ref{f1}(c), a low-energy ABS will be trapped in the QD before the topological phase transition point. In contrast to the MZMs that distribute non-locally at both ends [Fig. \ref{f1}(a)], the ABS is a bound state localized at one end of the nanowire [Fig. \ref{f1}(b)]. By replacing a pair of MZMs with an ABS in the modified ``tetron'', distinctly different braiding result will be obtained.
The final states after the same braiding process are time-dependent and oscillate with the braiding time-cost $T$, which is in stark contrast to the $T$-independent behavior of the braiding with true MZMs. As shown in Fig. \ref{f2}(a)-(c), $\psi_1^{-} (6T)$ is equal to $\psi_1^{+} (0)$ at $T = 400/\Delta$, and turns into a superposition of $\psi_1^{\pm}$ and $\psi_2^{\pm} $ at $T = 450/\Delta$, and then comes back to $\psi_1^{+} (0)$ at $T = 500/\Delta$.

\begin{figure}
\centering
\includegraphics[width=3.25in]{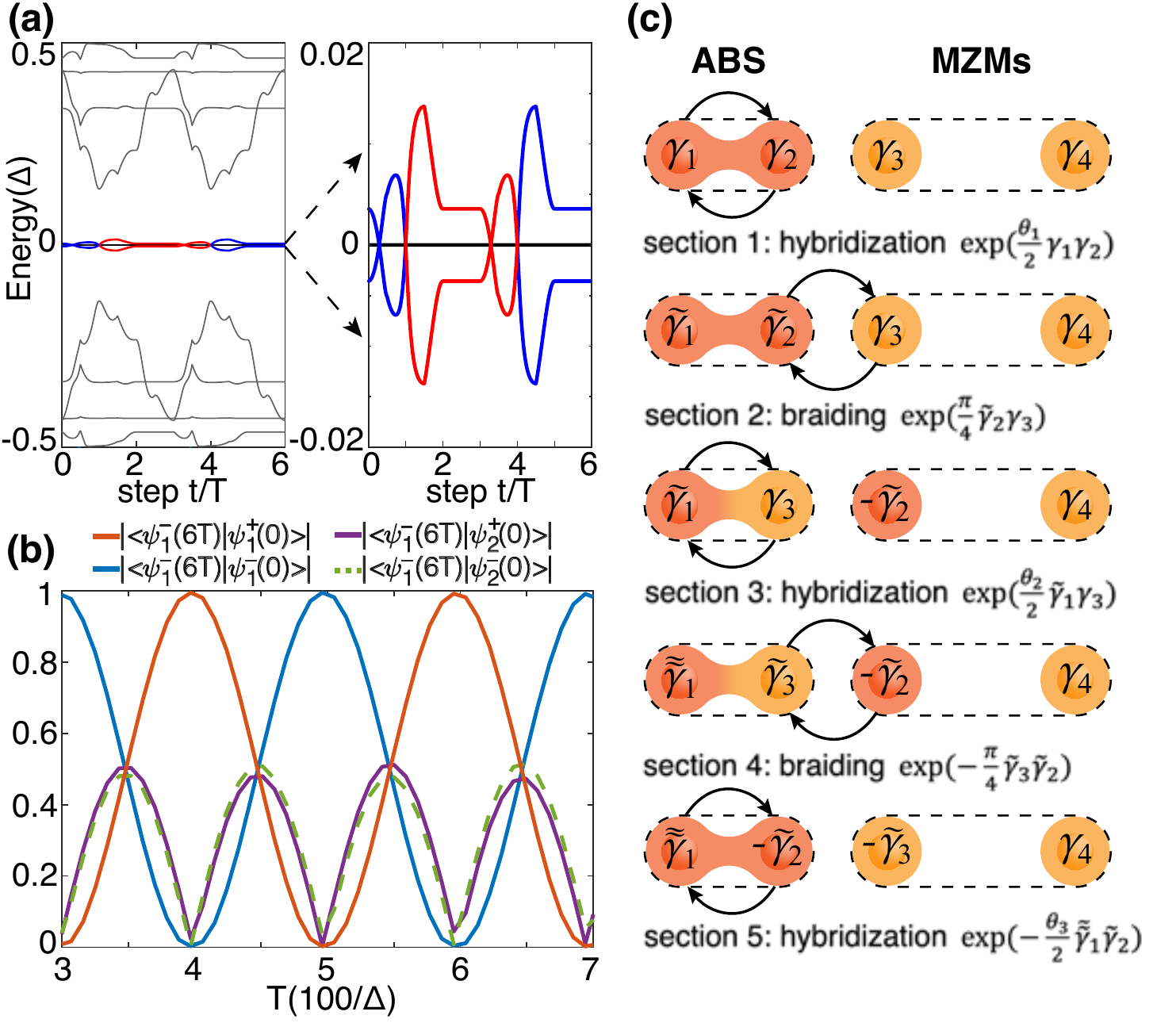}
\caption{
(a) The eigenenergy of the ABS during the braiding.
(b) Braiding results as functions of the braiding time-cost $T$.
(c) An illustration of the braiding process in the presence of an ABS. The whole braiding operation swapping $\gamma_2$ and $\gamma_3$ twice composes of five sections: section 1, 3, and 5 are dynamical evolution caused by the nonvanishing hybridization energy; section 2 and 4 are non-Abelian braiding between two MZMs.}
\label{f3}
\end{figure}

The local density of states (LDOS) distribution in Fig. \ref{f2}(d)-(i) unveils the temporal and spatial profile for both the ABS and the MZMs during the braiding. At the beginning, the LDOS of ABS exhibits a twin-peak structure, implying that the ABS can be treated as a pair of coupled MZMs $\gamma_1$ and $\gamma_2$ which are spatially separated with a finite distance. Such results are consistent with the previous studies \cite{ChunXiao1, Wimmer, Aguado2}.
In this point of view, the braiding in the presence of an ABS could be equivalent to the exchange between one ``free'' MZM and another MZM bounded in the ABS \cite{SN1}. Specifically, two additional process in addition to the braiding will happen as indicated by the LDOS:
bounded MZM $\gamma_2$ being moved out of ABS (splitting) in step 1, and free MZM $\gamma_3$ being moved into the ABS (fusion, reverse of the splitting) in step 2.
During the first step ($t\in[0,T]$), the hybridization energy $\epsilon_{1}(t)$ between the two MZMs $\gamma_1$ and $\gamma_2$ is relatively large due to the splitting process, hence it cannot be neglected as demonstrated in Fig. \ref{f3}(a) \cite{SN2}.
The Hamiltonian for hybridization is $H_{\mathrm{eff}}(t) = i \epsilon_{1}(t) \gamma_{1} \gamma_{2}$ and the corresponding evolution operator is $U_1=e^{\frac{\theta_1}{2}\gamma_1\gamma_2}$, where $\frac{\theta_1}{2}=\int_0^{T} \epsilon_{1}(t) \mathrm{d}t$ is the phase accumulated during the evolution. Such hybridization is equivalent to a unitary transformation on $\gamma_1$ and $\gamma_2$ \cite{Sun}:
\begin{eqnarray}
\label{model2}
\tilde{\gamma}_1 &=& U_1^{\dagger}\gamma_1U_1 = \cos(\theta_{1})\gamma_1+\sin(\theta_{1})\gamma_2 \nonumber, \\
\tilde{\gamma}_2 &=& U_1^{\dagger}\gamma_2U_1 = -\sin(\theta_{1})\gamma_1+\cos(\theta_{1})\gamma_2.
\end{eqnarray}
\noindent At time $t=T$, $\tilde{\gamma}_2$ is fully separated from $\tilde{\gamma}_1$. After that, $\gamma_3$ starts to move into the ABS and hybridize with $\tilde{\gamma}_1$. In the meantime, $\tilde{\gamma}_2$ and $\gamma_3$ are swapped and the corresponding braiding operator is $B(\tilde{\gamma}_2,\gamma_3) = e^{\frac{\pi}{4}\tilde{\gamma}_2\gamma_3}$.

The whole braiding operation can be decomposed into five braiding sections as sketched in Fig. \ref{f3}(c). The evolution operator for such whole braiding process can be expressed as the product of the evolution operator in each of these five steps as $U_F = e^{\frac{\theta_1}{2}\gamma_1\gamma_2}
e^{\frac{\pi}{4}\tilde{\gamma}_2\gamma_3}
e^{\frac{\theta_2}{2}\tilde{\gamma}_1\gamma_3}
e^{\frac{\pi}{4}\tilde{\gamma}_2\tilde{\gamma}_3}
e^{-\frac{\theta_3}{2}\tilde{\tilde{\gamma}}_1\tilde{\gamma}_2}$, where $\tilde{\tilde{\gamma}}_1 =
e^{-\frac{\theta_2}{2}\tilde{\gamma}_1\gamma_3}\tilde{\gamma}_1
e^{\frac{\theta_2}{2}\tilde{\gamma}_1\gamma_3}$,
$\tilde{\gamma}_3 = e^{-\frac{\theta_2}{2}\tilde{\gamma}_1\gamma_3}\gamma_3
e^{\frac{\theta_2}{2}\tilde{\gamma}_1\gamma_3}$,
$\frac{\theta_2}{2}=\int_T^{4T} \epsilon_{2}(t) \mathrm{d}t$, and
$\frac{\theta_3}{2}=\int_{4T}^{6T} \epsilon_{3}(t) \mathrm{d}t$ \cite{SN3}.
Therefore, in contrast to the MZMs' non-Abelian braiding which merely depends on the topology, the braiding in the presence of ABS is process-dependent.
Such braiding exhibits accumulative behavior that the evolution in each step relies on the evolution result in the previous step. For instance, the state $\psi_1^{-}(0) = (\gamma_{1}-i\gamma_{2}) / \sqrt{2}$ evolve into $\psi_1^{-}(6T) = (\tilde{\tilde{\tilde{\gamma}}}_1+i\tilde{\tilde{\gamma}}_2) / \sqrt{2}$, in which both $\tilde{\tilde{\tilde{\gamma}}}_1=e^{\frac{\theta_3}{2}\tilde{\tilde{\gamma}}_1\tilde{\gamma}_2}\tilde{\tilde{\gamma}}_1
e^{\frac{\theta_3}{2}\tilde{\tilde{\gamma}}_1\tilde{\gamma}_2}$ and $\tilde{\tilde{\gamma}}_2=e^{\frac{\theta_3}{2}\tilde{\tilde{\gamma}}_1\tilde{\gamma}_2}\tilde{\gamma}_2
e^{-\frac{\theta_3}{2}\tilde{\tilde{\gamma}}_1\tilde{\gamma}_2}$ depends on the evolution in the previous step (the other states also show similar behaviors, see \cite{S1}). Hence, the weight of $\psi_1^{-} (6T)$ on $\psi_1^{\mp}(0)$ oscillates in a sinusoidal behavior as $[1\mp\cos(\theta_2)]/2$, and on $\psi_2^{\pm}(0)$ oscillates as $\sin(\theta_2)/2$.
The numerical results shown in Fig. \ref{f3}(b) is fully consistent with the analytical prediction.
 Moreover, according to the peak positions and the oscillation period of $|\langle \psi_1^{-}(6T) | \psi_1^{\pm}(0) \rangle|$ [see Fig. \ref{f3}(b)], it can be concluded that $\theta_2 = 2\pi T/\Delta T$.
Remarkably, the explicit form of $\theta_2$ suggests that the non-Abelian braiding properties of the MZMs can still be exhibited even when an ABS is involved. Specifically, $\theta_2=2\pi T/\Delta T$ indicates that $\psi_1^-(6T)$ will evolve into $\psi_1^+(0)$ if the dynamic phase can be eliminated, implying the presence of a geometric phase of $\pi$. In other words, if the geometric phase is absent, then the explicit form of $\theta_2$ should be $\theta_2 = 2\pi (T/\Delta T+1/2)$.

{\emph {Manipulating the braiding results}} --- In the many-body basis $(|0\rangle, \Psi_1^{\dagger}|0\rangle,\Psi_2^{\dagger}|0\rangle,\Psi_1^{\dagger}\Psi_2^{\dagger}|0\rangle)$, $U_F$
has the matrix form of

\begin{equation}
 i\begin{bmatrix}
   e^{i\frac{\theta_1+\theta_3}{2}}S_{\frac{\theta_2}{2}} & 0 & 0 &e^{i\frac{\theta_1-\theta_3}{2}}C_{\frac{\theta_2}{2}}\\
   0 & -e^{-i\frac{\theta_1+\theta_3}{2}}S_{\frac{\theta_2}{2}} & e^{-i\frac{\theta_1-\theta_3}{2}}C_{\frac{\theta_2}{2}} & 0\\
   0 & e^{i\frac{\theta_1-\theta_3}{2}}C_{\frac{\theta_2}{2}} & -e^{i\frac{\theta_1+\theta_3}{2}}S_{\frac{\theta_2}{2}} & 0\\
   e^{-i\frac{\theta_1-\theta_3}{2}}C_{\frac{\theta_2}{2}} & 0 & 0 &e^{-i\frac{\theta_1+\theta_3}{2}}S_{\frac{\theta_2}{2}} \\
  \end{bmatrix}.
\end{equation}

\noindent where $S_{\frac{\theta_2}{2}} \equiv \sin(\theta_2/2)$ and $C_{\frac{\theta_2}{2}} \equiv \cos(\theta_2/2)$. This off-diagonal form of the braiding matrix demonstrates the non-Abelian nature of the braiding operation. Moreover, by tuning $\theta_2$ and $\theta_3$, the corresponding qubit can be rotated into any point on the Bloch sphere by such braiding operation, which is beyond the MZM-based braiding operation.
Specifically, the polar angle $\theta_2$ is the dynamic phase accumulated during $t\in[T,4T]$, and the azimuthal angle $\theta_1 \pm \theta_3$ is determined by the period $t\in[0,T]$ and $t\in[4T,6T]$. Both these two phases can be modulated individually by tuning the parameters in the corresponding braiding sections.

\begin{figure}
\centering
\includegraphics[width=3.25in]{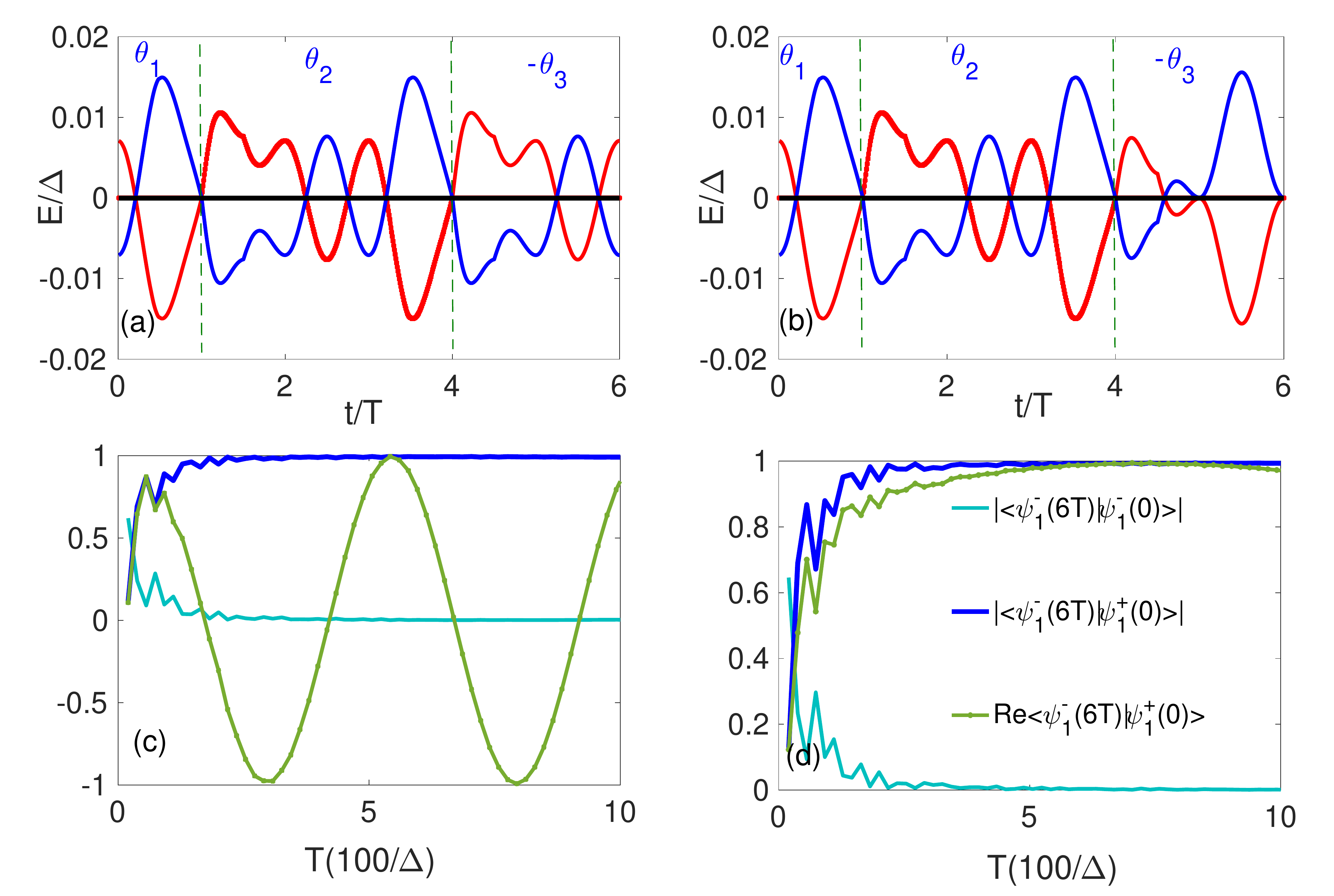}
\caption{
(a) Energy spectrum of the ABS in the presence of a sinusoidal magnetic field $V_x = [0.59+0.02\cos(t/T \cdot \pi)]\Delta$ during the braiding. The other parameters are the same as those in Fig. \ref{f3}. In this case, $\theta_2$ vanishes since the integration of the eigenenergy cancels out.
(b) Energy spectrum of the ABS with $V_x = [0.57+0.02\cos(t/T \cdot \pi)]\Delta$ during $t\in[4T,6T]$, the other parameters are the same as (a). In this case, both $\theta_2$ and $\theta_1-\theta_3$ vanishes.
(c) The braiding results for (a), where the amplitude of $\psi_1^-(6T)$ is independent of $T$, while the phase of $\psi_1^-(6T)$ oscillates with $T$.
(d) The braiding results for (b). Both the amplitude and the phase of $\psi_1^-(6T)$ are $T$-independent, which retrieves the original robust non-Abelian braiding properties.}
\label{f4}
\end{figure}

The manipulation of such phase angles can be assisted by 
combining the technology in geometric quantum computation (GQC)\cite{GQC1, GQC2, GQC3} which have also suggested by T. Karzig's group recently \cite{GQC4}. In GQC, the dynamic phase can be eliminated through the spin-echo technique which reverses the sign of the eigenenergy at the middle of the symmetric braiding protocol. Noticing that the spectrum of the ABS [Fig. \ref{f1}(c)] crosses the zero-energy in the vicinity of $V_x = 0.6\Delta$. Hence, it is possible to reverse the ABS's eigenenergy by modulating the Zeeman energy.
For instance, by applying a sinusoidal magnetic field $V_x = [V_{x0}+V_{x1}\cos(t/T \cdot \pi)]$, the eigenenergy will cross the zero-energy during the braiding. In such situation, the oscillation period will increase tenfold since the dynamic phase is nearly eliminated. In the special case of $V_{x0}=0.59\Delta$ and $V_{x1} =0.02\Delta$, as shown in Fig. \ref{f4}(a), the dynamic phase $\theta_2$ can be completely canceled out (Actually, $\theta_2$ is also canceled out if $V_{x1}=0.01\Delta$, indicating $V_{x1}$ does not need fine-tuning), so that the braiding result $\psi_{j}^{+} \to \psi_{j}^{-}$ independent of the braiding time-cost $T$ retrieves [Fig. \ref{f4}(c)]. However, since the azimuthal angle $\theta_1-\theta_3$ does not vanish, $\mathrm{Re} \langle\psi_{1}^{+}(6T)|\psi_{1}^{-}(0)\rangle $ still oscillates with $T$.
Considering the fact that the modulation could be individually performed in each braiding section, by tuning $V_{x0} = 0.570\Delta$ during $t\in[4T,6T]$ while keeping the other parameters invariant, $\theta_1-\theta_3$ can also be eliminated. Therefore, the non-Abelian braiding recovers the $T$-independent form [Fig. \ref{f4}(d)] as in the case that only MZMs are involved.


{\emph {Discussion}} ---
We have shown that the non-Abelian braiding of MZMs can be well performed in a modified ``tetron'' structure. Furthermore, we also investigate the MZMs' braiding properties when a low-energy ABS is presented. Although we have not discussed the noise effect induced by the ABS \cite{Nois1,Nois2}, we can rationally expect that its influence is quite small, since the previous studies indicate that the spin-echo techniques can largely reduce the noise effect \cite{GQC4}.
Finally, we want to point out that the phase elimination method discussed above can also be performed for the finite-size-effect-induced partially overlapped MZMs since the ABS is deemed as a pair of weakly coupled MZMs with finite separation. In one of our previous works \cite{ABS2}, we have revealed that the spectrum of such partially overlapped MZMs will cross at zero-energy with definite parity by modulating the Zeeman field or the gate voltage. It implies that the dynamic evolution can be well manipulated by modulating either the Zeeman field or the gate voltage. Therefore, the TQC can be realized in a shorter TSC nanowire, which possesses advantages such as supporting universal gate operation through modulating the dynamical evolution.

{\emph{Acknowledgement}} --- Wenqin Chen and Jiachen Wang contributed equally to this work. This work is financially supported by NSFC (Grants No. 11974271) and NBRPC (Grants No. 2017YFA0303301, and No. 2019YFA0308403).


\begin{thebibliography}{WSM_JJ}
\bibitem{SN2} Even though the intial energy of ABS is very small, the hybridization energy induced by the splitting and fusion process of MZMs are usually one the order of $10^-{2}\Delta$.
\bibitem{SN3} Here $\epsilon_{2}$ is induced by the hybridization between $\tilde{\gamma}_1$ and $\gamma_3$, while $\epsilon_{3}$ is induced by the hybridization between $\tilde{\tilde{\gamma}}_1$ and $-\tilde{\gamma_2}$.
\end{thebibliography}
\end{document}